\documentclass[a4paper,11pt]{article}
\usepackage{pos}

\usepackage{xcolor}
\usepackage{amsmath, braket, mathtools}
\usepackage{dsfont}
\usepackage{float}
\usepackage{subcaption}

\usepackage{hyperref}
\usepackage{cleveref}

\title{Testing the $\mathrm{SU}(2)$ lattice Hamiltonian built from $S_3$ partitionings}

\author*[a]{Marco Garofalo}
\author[b,c]{Tobias Hartung}
\author[a]{Timo Jakobs}
\author[d, e]{Karl Jansen}
\author[f]{Johann Ostmeyer}
\author[a]{Dominik Rolfes}
\author*[a]{Simone Romiti}
\author[a]{Carsten Urbach}

\affiliation[a]{HISKP (Theory), Rheinische Friedrich-Wilhelms-Universität Bonn, Nußallee 14-16, D-53115 Bonn, Germany}
\affiliation[b]{Northeastern University - London, Devon House, St Katharine Docks, London, E1W 1LP, UK}
\affiliation[c]{Khoury College of Computer Sciences, Northeastern University, 440 Huntington Avenue, 202 West Village H, Boston, MA 02115, USA}
\affiliation[d]{CQTA, DESY Zeuthen, Platanenallee 6, 15738 Zeuthen, Germany}
\affiliation[e]{Computation-Based Science and Technology Research Center,
The Cyprus Institute, 20 Kavafi Street, 2121 Nicosia, Cyprus}
\affiliation[f]{Department of Mathematical Sciences, University of Liverpool, Liverpool, L69 7ZL, United Kingdom}

\emailAdd{mgarofal@uni-bonn.de}
\emailAdd{simone.romiti@uni-bonn.de}

\abstract{ We test a possible digitization of SU$(2)$ lattice gauge theories 
based on partitionings of the sphere $S_3$.
In our construction the link operators are unitary and diagonal, 
with eigenvalues determined by the vertices of the partitioning.
The canonical momenta are finite difference operators approximating 
the Lie derivatives on the manifold.
In this formalism we implement the standard Wilson Hamiltonian.
We show results for a 2-site Schwinger-type model in 1D and a single-plaquette system in 2D.
Our calculations are performed on a classical computer, 
though in principle they can be implemented also on a quantum device. }

\FullConference{%
  The 40th International Symposium on Lattice Field Theory, LATTICE2023 \\
  31 July-4 August 2023 \\
  Fermilab, Batavia, Illinois, USA
}

\begin{document}
\maketitle

\section{Introduction}
\label{sec:intro}
The recent years have seen a revived interest in the Hamiltonian formulation of lattice gauge theories~\cite{dimeglio2023quantum,PhysRevD.101.114502,davoudi2022general,Zache:2023dko,Bauer:2023jvw}.
This is due to the rapid advancement of quantum computing resources. 
As an outstanding technological achievement, this could open the window for unprecedented efficiency for solving certain scientific problems.
One of them is certainly the quantum simulation of physical systems, namely the numerical solution of the time-dependent Schrödinger equation.
The goal is indeed to write down a truncated version of the Hilbert space and of the degrees of freedom of the theory, 
which recovers the exact version when the truncation is extrapolated to infinity.

The Hamiltonian formalism has in general many potential advantages compared to the classical Monte Carlo techniques based on the action~\cite{gattringer2009quantum}.
For instance, the time evolution of the physical states is performed in real time, hence not requiring careful analytic continuations.
There is no sign problem, one can simulate at non-zero baryon densities, and study dynamical events such as the scattering of bound states.
It is important to remark, however, that nowadays the available quantum devices still have a rather limited number of noisy qubits.
Therefore, many calculations are done on classical devices which are in practice limited to very small and simple systems.

The formulation of Hamiltonian gauge theories on the lattice has a long history~\cite{PhysRevD.11.395}, 
and can be treated in some cases with analytical techniques. 
Indeed, the seminal paper by Kogut and Susskind~\cite{PhysRevD.11.395} gave the formulation to infer the properties of an SU$(2)$ lattice gauge theory in the confining phase, 
in the approximation of strong coupling.
The formalism has been investigated and developed since then, also with the extension to generic SU$(N)$ symmetry~\cite{PhysRevD.91.054506}.
Eventually, the main goal for many particle physicists is SU$(3)$, as it is the gauge group of QCD.
%
The SU$(2)$ gauge theories represents nevertheless a topic of interest,
as they are cheaper to simulate and show a confining phase similar to QCD~\cite{PhysRevD.11.395,Sommer_1994}.

For the above reasons we have developed a digitization of an SU$(2)$ lattice gauge theory in the Hamiltonian formalism,
i.e. an algorithm for truncating the Hilbert space to a given size and building the matrices that approximate the degrees of freedom.
In the literature there is more than one choice for the basis of the Hilbert space, 
such as the \textit{electric}, \textit{magnetic}, \textit{loop-string-hadron}, etc.
Each basis and each prescription is expected to give accurate results in a given range of the Yang-Mills coupling constant.
In this paper we work in the magnetic basis, with unitary and diagonal links.

The Hilbert space is generated by the tensor product states for each lattice site and direction.
There we define the action of the operators $U$ (the link) and the $L_a,R_a$ (the canonical momenta)~\cite{Garofalo:2022swa,Jakobs:2023lpp,Romiti:2023hbd}.
The space these operators act on is isomorphic to a set of points on the SU$(2)$ group manifold, 
which we chart according to the isomorphism with the sphere $S_3$.
We call this set of points a \textit{partitioning}.
The eigenvalues of $U$ are SU$(2)$ color matrices corresponding to the points in the partitioning, 
while the canonical momenta are built as finite difference operators that approximate the Lie derivatives on the continuous manifold.
The fermionic fields are the usual ladder operators in the fermionic part of the Fock space.


\section{Theoretical background}
\label{sec:theor.bkg}
Here we outline the main features of the Hilbert space and degrees of freedom for a lattice $\mathrm{SU}(2)$ Hamiltonian.
These properties are standard in the literature. For a review, we defer to Refs.~\cite{PhysRevD.15.1128,PhysRevD.15.1111,PhysRevD.91.054506,PhysRevX.7.041046,Romiti:2023hbd}.

For a $\mathrm{SU}(2)$ lattice gauge theory the Hilbert space $\mathcal{H}$ is generated by the states:
\begin{equation}
  \label{eq:HilbertSpaceBasis}
  \ket{\mathcal{U}} \otimes \ket{{\chi}} =
  \bigotimes_{\vec{x}}
  \left[ \bigotimes_{\mu=1}^{d} \ket{\mathcal{U}_\mu(\vec{x})} \right] \,
  \otimes \ket{{\chi}(\vec{x})} \, .
\end{equation}
$\ket{\mathcal{U}}$ is the magnetic basis for the bosonic excitations, while $\ket{{\chi}}$ contains information about the fermions.
The $\mathcal{U}_\mu(\vec{x})$ are color matrices in the fundamental representation of $\mathrm{SU}(2)$,
generated by the normalized Pauli matrices $\tau_a = \frac{\sigma_a}{2}$.
The ${\chi}(\vec{x})$ are color spinors with entries given by the occupation number: either $0$ or $1$.
The corresponding operators $U_\mu(\vec{x})$ and $\chi(\vec{x})$ act as:
\begin{align}
  {U_\mu(\vec{x})^{ab} \ket{\mathcal{U}} \vcentcolon= \mathcal{U}_\mu(\vec{x})^{ab} \ket{\mathcal{U}}} \, , \\
  \label{eq:PsiFieldBehavior}
  {\chi^a}^\dagger \ket{\chi^a}(\vec{x}) = \ket{({\chi^a}(\vec{x}) + 1) \, \mathrm{mod} \, 2} \, \forall a \, .
\end{align}
The fermionic operators obey the anti-commutation relations:
\begin{equation} \label{eq:FermionsCommRel}
  \{\chi^a(\vec{x}), {\chi^b}^{\dagger}(\vec{y})\} = \delta_{ab} \delta_{\vec{x},\vec{y}} \, , \quad
  \{\chi^a(\vec{x}), {\chi^b}(\vec{y})\} = 0 \, .
\end{equation}
%

The $\mathrm{SU}(2)$ lattice Hamiltonian for a degenerate up-down quarks doublet coupled to gluons is
\begin{equation}
  \label{eq:OriginalKGHamiltonian}
  H = H_{\text{el}} + H_{\text{mag}} + H_F \, ,
\end{equation}
where:
\begin{align}
  H_{\text{el}}  & =
  \frac{g^{2}}{2}
  \sum_{\vec{x}} \sum_{\mu=1}^{d-1} \sum_{a=1}^{3}
  {(L_a)}_{\mu}^{2}(\vec{x}) \,                                              \\
  H_{\text{mag}} & =
  -
  \frac{1}{4 g^{2}}
  \sum_{\vec{x}} \sum_{\mu=1, \nu<\mu}^{d-1} \,
  \mathrm{Tr}[{U}_{\mu \nu}(\vec{x}) + {U}_{\mu \nu}^\dagger(\vec{x}) ] \, , \\
  H_F            & =
  \mu \sum_{\vec{x}} (-1)^{\vec{x}} {\chi^c}^\dagger(\vec{x}) \chi^c(\vec{x})
  + \sum_{\vec{x}} \sum_{\mu=1}^{d-1}{\chi^c}^\dagger(\vec{x}) U_\mu(\vec{x})^{c c'} \chi^{c'}(\vec{x}+\mu)
  \, .
\end{align}
$\vec{x}$ is a generic site of the lattice and $d$ is the number of dimensions.
$H_{\text{el}}$, $H_{\text{mag}}$, $H_F$ are the electric, magnetic and fermionic part of the Hamiltonian respectively.
${U}_{\mu \nu}(\vec{x})$ is the plaquette operator and the trace is taken in color space only.
The Yang-Mills coupling $g$ is related to the usual coupling $\beta$ by $\beta = 4 / g^2$.

The system is gauge invariant under gauge transformations generated by the 
Left and Right canonical momenta $L_a$ and $R_a$.
In analogy to electrodynamics, these are often called \textit{electric} fields
 (more detail in Section.~\ref{sec:momenta}).

The gauge redundancy of the theory is removed by restricting to the subspace of $\mathcal{H}$ of physical states,
defined by \textit{Gauss' law}:
\begin{equation}
  \label{eq:GaussLawSpatialCondition}
  G_a(\vec{x}) \ket{\psi}_{\text{phys.}} =
  \left[ \sum_{\mu=1}^{d} {(L_a)}_\mu(\vec{x}) + {(R_a)}_\mu(\vec{x}-\mu) - Q_a(\vec{x}) \right] \ket{\psi}_{\text{phys.}}
  = 0
  \,\, ,
\end{equation}
where $Q$ is the charge operator:
\begin{equation} \label{eq:ChargeOperator}
  Q_a = \frac{1}{2} \chi^\dagger(\vec{x}) \,  \tau^a \, \chi(\vec{x}) \, .
\end{equation}

We conclude by recalling that despite the simplicity of the fermionic Fock space,
the gauge part has a rich structure already by itself.
In fact, a basis for physical states for the pure gauge theory is isomorphic to the irreps of the $\mathrm{SU}(2)$ algebra.
The latter are labelled by 3 half-integers $j, m_L, m_R$, with the constraint $|m_L|, |m_R| = -j,\ldots,j$.
$j$ is the main quantum number, while $m_L$ and $m_R$ are the left and right magnetic quantum numbers respectively.
When we truncate the Hilbert space, we cannot preserve the unitarity of the links and the exact canonical commutation relations simultaneously.
In this work we keep the former, sampling the eigenvalues of $U$ from a discrete set points on $S_3$ (see upcoming sections for more details).

\section{Construction of the momenta}
\label{sec:momenta}
The main difficulty in our prescription is the construction of the links and momenta.
We do this by  using the isomorphism of the $\mathrm{SU}(2)$ group manifold with the sphere $S_3$ in $4$ dimensions.
The latter is partitioned into a finite number of tetrahedra, with $N$ vertices labelled by three angles $\vec{\alpha}$. 
The link $U$ is diagonal on the Hilbert space, with eigenvalues given by $2\times 2$ color matrices $\mathcal{U}$:
\begin{equation}
U \ket{\vec{\alpha}} = \mathcal{U(\vec{\alpha})} \ket{\vec{\alpha}} \, .
\end{equation}
The convention for the chart map $\vec{\alpha}$ defines the form of $\mathcal{U(\vec{\alpha})}$ in terms of the generators $\tau_a$.
The canonical momenta are constructed as the finite difference operators approximating the Lie derivatives along the directions of the generators:
\begin{align}
  {L}_{a} f(U) & = -i \frac{\mathrm{d}}{\mathrm{d} \omega} f( e^{-i \omega \tau_a} U) |_{\omega=0} \, ,\\
  {R}_{a} f(U) & = -i \frac{\mathrm{d}}{\mathrm{d} \omega} f( U e^{i \omega \tau_a}) |_{\omega=0} \, ,
\end{align}
where $f$ is an arbitrary $L^2$-integrable function over the $\mathrm{SU}(2)$ Haar measure.

In \cite{Garofalo:2022swa,Jakobs:2023lpp} this was done using a finite element method on a Delaunay triangulation of $S_3$,
and the derivatives are obtained as average over all the simplices of the 1st order finite differences.
The momenta are ultra-local, namely the non-vanishing matrix elements are either on the diagonal or couple nearest neighbors on $S_3$.
This construction was implemented for several types of partitionings of $S_3$ (linear, Fibonacci, etc.) 
becoming dense in the limit of infinite number of points $N$.
In that work we showed numerical evidence for power law convergence in $N$ to the correct Lie algebra structure and
canonical commutation relations, by looking at the action of these operators on test functions.
We also showed the convergence of the spectrum of the electric part of the Hamiltonian to the continuum manifold one.
For the latter we employ the operator $\sum_a L_a L_a$ found directly as the approximation of the Laplace-Beltrami operator on $S_3$,
while the 1st order finite differences $L_a, R_a$ are used for the construction of the Gauss' law generators.
This requires some care, as the latter contain spurious effects coming from the discretization of the manifold.
In the following we will refer to this type of construction by the name of the partitioning of $S_3$ employed.
For details on the different constructions see Ref.~\cite{Jakobs:2023lpp}.

In Ref.~\cite{Romiti:2023hbd}, we have provided a construction that is exact on a subspace of the Hilbert space,
while the remaining subspace can be projected to arbitrary energies above the cutoff.
This property causes a loss of locality of the matrices in the manifold space, 
while the operators are still local on the space lattice.
Thus, this construction is expected to work up to some maximum value for the main quantum number $j$.
In the following we will refer to it as DJT, 
as the canonical momenta are constructed using what in Ref.~\cite{Romiti:2023hbd} is called Discrete Jacobi Transform.

\section{Construction of the Hamiltonian} 
\label{sec:construction}
We generate the Hilbert space according to a truncated version of \cref{eq:HilbertSpaceBasis}.
Hence, it will be a finite dimensional vector space, 
with the basis given by the tensor products of the configuration of the degrees of freedom.

At every point, the algebra of the degrees of freedom is the same.
Therefore, it is sufficient to define their action ``pointwise'',
so that the operators will simply be tensor products of the identity at all points but for the one we consider.
More explicitly, if we have an operator $\mathcal{O}(x,\mu)$ then:
%
\begin{equation}
    \label{eq:OperatorsLatticeTensorProduct}
    \mathcal{O}(x,\mu) = 
    \left[\bigotimes_{i < i_{x, \mu}} \mathds{1} \right] 
    \otimes \mathcal{O} \otimes
    \left[ \bigotimes_{i > i_{x, \mu}} \mathds{1} \right]
    \, ,
\end{equation}
where $i_{x, \mu}$ is the checkerboard index corresponding to the point of the space lattice.
Furthermore, the fundamental degrees of freedom are factorizable into the gauge (G) and fermionic (F) part, 
so that each $\mathcal{O}$ is either ${\mathcal{O}_G \otimes \mathds{1}_F}$ or ${\mathds{1}_G \otimes \mathcal{O}_F}$. 

In our case, the gauge operators are the link $U$ and the canonical momenta ${L_a, R_a}$.
The fermionic ones are the $\chi$s.
In this work we considered at most one light quark doublet, 
but the formalism is the identical for multiple ones.
We remark that at finite volume (i.e.~finite number of space lattice points), 
the fermionic Hilbert space is always finite.
Therefore, the field $\chi$ at every point $\vec{x}$ can be simply implemented according to \cref{eq:PsiFieldBehavior}.

The Hamiltonian is built according to
\cref{eq:OriginalKGHamiltonian,eq:OperatorsLatticeTensorProduct},
and Gauss' law is enforced by adding a penalty term:
\begin{equation}
H_{\text{pen}} = \kappa \sum_{\vec{x}, \mu} \sum_a (G_a)_\mu(\vec{x}) (G_a)_\mu(\vec{x}) \, ,
\end{equation}
where $\kappa > 0$.
As $\kappa$ increases, so do the energies of the unphysical states.
According to the decoupling theorem, these modes decouple more and more from the theory.
The advantage of this approach is the simplicity of the implementation, 
as it doesn't require a maximal tree construction beforehand.
On the lattice $\kappa$ has to be chosen to be above the cutoff $\sim 1/a$ induced by lattice spacing $a$.
In practice, find the optimal value of $\kappa$ numerically: 
it has to be large enough such that the results are independent of a further increase.
In the case of the momenta obtained with the finite element method,
some extra care is required. In fact that prescription possesses intrinsic spurious effects, 
coming from the discretization of the manifold.
Therefore, $\kappa$ can't be too large in order to not induce large residual discretization effects.

\section{Numerical results}
\label{sec:numerical.results}

In this section we show our numerical results obtained with the constructions described in \cref{sec:construction}.
These are obtained on a classical computer by an exact diagonalization of the Hamiltonian.

\subsection{Schwinger model in 1D}
\label{sec:results.schwinger}

We consider the Hamiltonian for a 2-site lattice with open boundary conditions.
Written explicitly, the Hamiltonian reads:
\begin{equation} \label{eq:HSchwinger1D}
    H = \mu \sum_{k=1}^{N_{\textrm{sites}}}
    \sum_{c=1}^{2} (-1)^k \, \chi^{c \dagger}_k \chi^c_k + \sum_{k=1}^{N_{\text{links}}} ({L}^a_k)^2 + x \sum_{k=1}^{N_{\textrm{links}}} \sum_{c,c'=1}^{2} \left( \chi^{c \, \dagger}_k {U}_k^{cc'}
    \chi^{c'}_{k + 1}  + \textrm{h.c.} \right) \, ,
\end{equation}
where we have just rewritten \cref{eq:OriginalKGHamiltonian} for this particular system.
The fermionic $\chi$ operators obey the commutation relations of \cref{eq:FermionsCommRel}.

$H$ describes the dynamics of a Schwinger-type model in one
spatial dimension (1D) with SU$(2)$ gauge symmetry,
on a lattice connecting two sites.
Since there is no plaquette term in 1D, $H_{\text{mag}}=0$
and Gauss' law can be enforced exactly by integrating out
the gauge fields analytically. 
The resulting Hamiltonian reads~\cite{PhysRevX.7.041046}:
\begin{equation}\label{eq:HSchwinger1DIntegrated}
    \begin{split}
        {H}_{\textrm{integrated}} &= x \sum_k \left( \chi^{c \,
            \dagger}_k	\chi^{c}_{k + 1}	+ \textrm{h.c.} \right)
         + \mu \sum_{k=1}^{N_{\textrm{sites}}} \sum_{c=1}^{2}
        (-1)^k \,
        \chi^{c \dagger}_k \chi^c_k + \sum_{a=1}^3 \sum_{k, k' = 1}^{N_{\textrm{sites}}} Q_k^a
        \,
        \frac{|k-k'|}{2} \, Q_{k'}^a \, ,
    \end{split}
\end{equation}
where $Q_a$ is the charge operator defined in \cref{eq:ChargeOperator}.

\begin{figure}[H]
    \centering
    \includegraphics[width=1.0\textwidth]{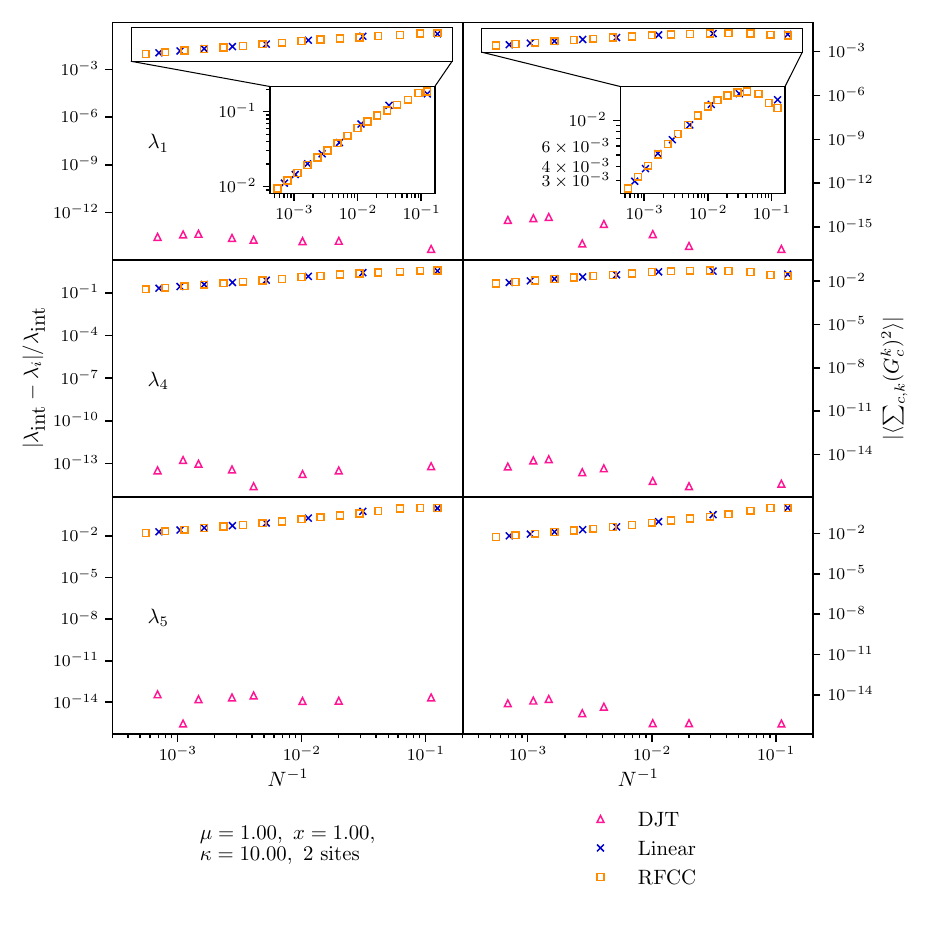}
    \caption{
        Spectrum of the 1D Schwinger-type model with two lattice sites. 
        We show the (normalized) difference between the eigenvalues of the
        ``standard'' lattice Hamiltonian  of \cref{eq:HSchwinger1D} 
        and the analytical results obtained from
        \cref{eq:HSchwinger1DIntegrated} as a function of $1/N$ for
        $\lambda_{1,4,5}$ in the left column. In the right column the
        residual Gauss' law violations are show, again as a function
        of $1/N$ for the respective three eigenstates. We compare
        Linear and RFCC partitionings to the DJT construction.
    }
    \label{fig:SchwingerSpectrumWithDJT}
\end{figure}

We check the convergence of results
obtained with Hamiltonian from \cref{eq:HSchwinger1D} to the
exact results obtained from ${H}_{\textrm{integrated}}$ numerically. 
In \cref{fig:SchwingerSpectrumWithDJT} we show in the left column the
eigenvalue deviations
$(\lambda_\mathrm{int}-\lambda_i)/\lambda_\mathrm{int}$ for $i=1$,
$i=4$ and $i=5$ as a function of the number of points $N$ in the
respective partitionings for $\mu=1, x=1$ and $\kappa=10$. In the
right column the residual Gauss' law violations for the respective
eigenstates are plotted. 
We compare DJT with the linear and the RFCC partitionings.

The eigenvalues $\lambda_2$ and $\lambda_3$ are omitted, as they
manifestly vanish for any number of points of the partitioning.
In fact, the canonical momenta are finite difference operators with
eigenvalue zero on the $\ket{j=0}$ state.  Moreover, the state
$\ket{00} \otimes \ket{j=0} \otimes \ket{00}$ (or the same with
$\ket{00} \to \ket{11}$) is an eigenstate with eigenvalue zero for
both Hamiltonians $H$ and ${H}_{\textrm{integrated}}$. This is a
feature of the model and of the peculiar 2-site lattice
configuration.

We observe convergence in $1/N$ for both, linear and RFCC
partitionings, see also the inlets in the upper row. For DJT, however,
the eigenvalues agree with the analytical results up to machine
precision and the residual Gauss' law violations are also zero up to
round off errors.

We conclude that in this case the DJT is advantageous when compared to
the Linear and RFCC partitionings. It is likely
a feature of this particular system, as with only two sites Gauss' law
forbids heavier states.

\subsection{Single plaquette system}
\label{sec:single.plaquette}

Here we present our preliminary results for the single plaquette system.
This is a $4$-sites system in $2$-dimensions, whose Hamiltonian is:
\begin{equation} \label{eq:HPlaquette2D}
    \hat{H} = \frac{g^2}{2}\sum_{c=1}^3\sum_{i=0}^3 (\hat{L}^c_i)^2 -
    \frac{2}{g^2}
    \mathrm{Tr}
    \left[ \hat{U}_0 \hat{U}_1 \hat{U}_2
        \hat{U}_3\right] \, .
\end{equation}
The latter is \cref{eq:OriginalKGHamiltonian}, without fermions, for this particular setup.
Gauss' law reads:
\begin{equation}
    \left(\hat{L}^c_i + \hat{R}^c_{(i+1) \,\, \mathrm{mod} \,\, 4}\right)
    \ket{\psi}_{\text{phys.}} = 0 \, .
\end{equation}

In order to ease the numerical simulation, 
we enforce the spacetime symmetries analytically, thus reducing the dimension of the relevant Hilbert space.
We write the gauge part of the state as $\ket{\ell_0,\ell_1,\ell_2,\ell_3}$, 
where $\ell_i$ is the configuration of the $i$-th link.
The system is symmetric under spatial rotations of $\pi/2$ degrees, 
which form a discrete group $Z_4$ with generator $\hat{r}$ acting as:
\begin{equation}
    \hat{r} \ket{\ell_0,\ell_1,\ell_2,\ell_3} = \ket{\ell_3,\ell_0,\ell_1,\ell_2} 
    \, .
\end{equation}
The eigenstates of $\hat{r}$ are classified by a number $n$ as follows:
\begin{equation}
    \ket{\phi_{\hat{r}}}_n \sim \left(1 + \lambda_{\hat{r}}^n \hat{r} +
    (\lambda_{\hat{r}}^n)^2
    \hat{r}^2 + (\lambda_{\hat{r}}^n)^3 \hat{r}^3 \,\right)
    \ket{\ell_0,\ell_1,\ell_2,\ell_3} \, ,
\end{equation}
where $\lambda_n = e^{ i n \pi/2}$, $n=0,\ldots,3$.
Note that the eigenstates $\ket{\phi_{\hat{r}}}_n$ as defined above vanish for
some combinations of $(\ell_0,\ell_1,\ell_2,\ell_3)$. 
In particular states where $\ell_0 = \ell_2$ and $\ell_1 = \ell_3$
will only give valid eigenstates for $n=0,2$. 
For $\ell_0 = \ell_1 = \ell_2 = \ell_3$ only $n=0$ gives a non-vanishing eigenstate. 
With this, we find numerically that the ground state and first excited state belong to in the $n=0$ subspace. 
Thus, it is sufficient to only diagonalize this smaller block.

\begin{figure}
    \centering
    \includegraphics[width=1.0\textwidth]{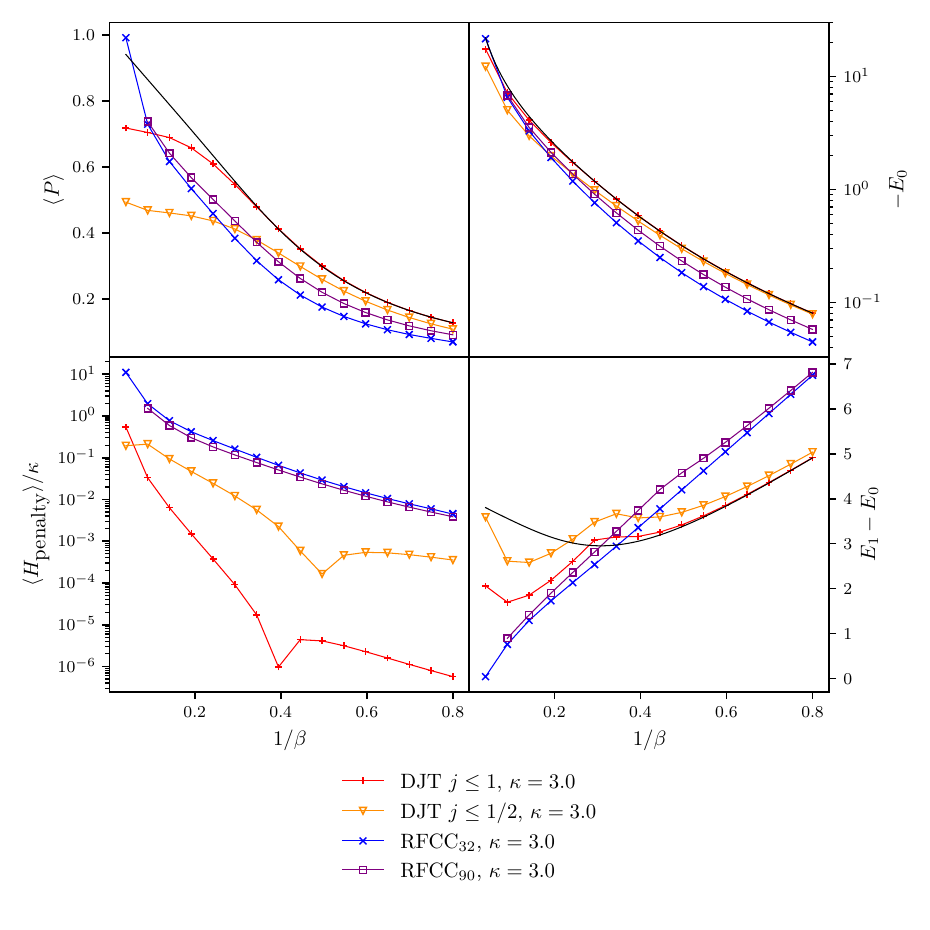}
    \caption{
      Expectation value of the plaquette $\langle P\rangle$, of the
      vacuum energy $E_0$, the gap $E_1-E_0$, and $\langle
      H_\mathrm{penalty}\rangle/\kappa$ as a function of
      $1/\beta=g^2/4$. We compare variational results (solid lines) to
      results obtained with two particular RFCC partitionings with
      $N=32$ and $N=90$ and with two DJT constructions for $j\leq1/2$ and
      $j\leq 1$ with $N=9$ and $N=50$, respectively. 
    }
    \label{fig:results.plaquette}
\end{figure}

We compute the expectation value of the plaquette in the numerically found vacuum state,
together with the mass gap of the theory, for a range of couplings $\beta$. 
In \cref{fig:results.plaquette} we compare our preliminary results to
the ones obtained through a variational ansatz~\cite{HUANG1988733}: the solid
black lines represent the variational results, and the points the
results from exact diagonalization of the digitized Hamiltonians. With
the RFCC partitioning we observe that with increasing number of points
in the partitioning the results for $\langle P\rangle$
approach the variational result for $1/\beta\lesssim 0.3$, see the upper
left panel. For the ground state energy this is even true for the full
range of $\beta$-values considered, see the upper right panel. For the
gap $E_1-E_0$, on the other hand, this approach is less visible, at
least for the number of points in the partitionings investigated here.

For the DJT construction, shown as (blue) crosses and (purple) boxes,
we observe smaller deviations from the variational result than with
the aforementioned RFCC partitionings for $1/\beta\lesssim 0.3$ for all
three quantities. Also $\langle H_\mathrm{penalty}\rangle/\kappa$ -- a
measure for residual violations of Gauss' law in the vacuum -- is
smaller for the DJT construction compared with the RFCC
partitionings. A the same time, the number of points in the respective
partitionings is smaller for DJT: for DJT with $j\leq 1/2$ 
$N = 9$ and for $j\leq 1$ $N = 50$ are required per
link, while the two shown RFCC partitionings do require $N=32$ and
$N=90$ points per link, respectively.

\section{Conclusion and outlook}
\label{sec:conclusion}
In this proceeding contribution, we have presented preliminary results
for different digitizations of a $\mathrm{SU}(2)$ lattice gauge theory
in the magnetic basis. We compare the constructions presented in
Refs.~\cite{Jakobs:2023lpp,Romiti:2023hbd}, based on the isomorphism
between $\mathrm{SU}(2)$ and the sphere $S_3$, which is sampled with a
finite number of points $N$ defining the truncated basis of the
Hilbert space. The gauge link operator is diagonal and unitary, while
the canonical momenta are finite difference matrices, converging to
the continuum manifold Lie derivatives at infinite truncation. 

Compared to the construction based on partitionings from
Ref.~\cite{Jakobs:2023lpp}, the one from Ref.~\cite{Romiti:2023hbd}
based on a discrete Jacobi transform (DJT) has the property that gauge symmetry
is exactly preserved on a subspace of the truncated Hilbert space for
the price of dense matrix representations of the canonical
momenta. Here we have tested and compared the different constructions
for two models, a Schwinger-type model with two sites and SU$(2)$
gauge symmetry, as well as a SU$(2)$ gauge theory consisting of a
single plaquette. For both models exact results are available to
compare with.

With the partitionings from Ref.~\cite{Jakobs:2023lpp} we observe
deviations from the exact results, which go to zero with increasing
number of points $N$ in the partitionings. With the
DJT~\cite{Romiti:2023hbd} on the other hand we find exact results up
to machine precision for the 2-site Schwinger-type model as a
consequence of the enhanced gauge symmetry. While not exact anymore,
DJT shows also for the single plaquette gauge theory smaller
deviations from the expected result with smaller number of points in
the partitioning, at least for a certain range of
$\beta$-values when compared to the RFCC partitionings.

In the near future, we plan a systematical study of the single
plaquette gauge theory, both in parameter space of the theory as well
as the number of points of the partitioning.

\section*{Acknowledgements}
We would like to thank A.~Crippa for helpful discussions.
This work is supported by the Deutsche Forschungsgemeinschaft (DFG, German Research Foundation) and the  
NSFC through the funds provided to the Sino-German Collaborative Research Center CRC 110 “Symmetries
and the Emergence of Structure in QCD” (DFG Project-ID 196253076 - TRR 110, NSFC Grant No.~12070131001)
as well as by the STFC Consolidated Grant ST/T000988/1.
This work is supported with funds from the Ministry of
Science, Research and Culture of the State of Brandenburg within the Centre for Quantum Technologies and
Applications (CQTA). This work is funded by the European Union's Horizon Europe Framework Programme
(HORIZON) under the ERA Chair scheme with grant
agreement No. 101087126.

\begin{flushright}
    {\includegraphics[width=0.08\textwidth]{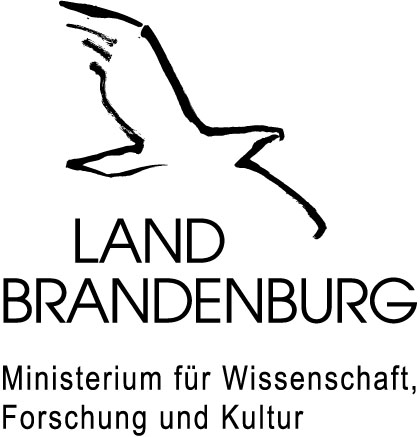}}
\end{flushright}

\bibliographystyle{jhep} 
\bibliography{refs_proceedings_2023} 

\providecommand{\href}[2]{#2}\begingroup\raggedright\begin{thebibliography}{10}

\bibitem{dimeglio2023quantum}
A.D.~Meglio, K.~Jansen, I.~Tavernelli, C.~Alexandrou, S.~Arunachalam,
  C.W.~Bauer et~al., \emph{Quantum computing for high-energy physics: State of
  the art and challenges. summary of the qc4hep working group},  2023.
\newblock 10.48550/arXiv.2307.03236.

\bibitem{PhysRevD.101.114502}
I.~Raychowdhury and J.R.~Stryker, \emph{Loop, string, and hadron dynamics in
  su(2) hamiltonian lattice gauge theories},
  \href{https://doi.org/10.1103/PhysRevD.101.114502}{\emph{Phys. Rev. D}
  {\bfseries 101} (2020) 114502}.

\bibitem{davoudi2022general}
Z.~Davoudi, A.F.~Shaw and J.R.~Stryker, \emph{General quantum algorithms for
  hamiltonian simulation with applications to a non-abelian lattice gauge
  theory}, \href{https://doi.org/10.48550/arXiv.2212.14030}{\emph{arXiv
  preprint arXiv:2212.14030} (2022) }.

\bibitem{Zache:2023dko}
T.V.~Zache, D.~Gonz\'alez-Cuadra and P.~Zoller, \emph{{Quantum and Classical
  Spin-Network Algorithms for q-Deformed Kogut-Susskind Gauge Theories}},
  \href{https://doi.org/10.1103/PhysRevLett.131.171902}{\emph{Phys. Rev. Lett.}
  {\bfseries 131} (2023) 171902}
  [\href{https://arxiv.org/abs/2304.02527}{{\ttfamily 2304.02527}}].

\bibitem{Bauer:2023jvw}
C.W.~Bauer, I.~D'Andrea, M.~Freytsis and D.M.~Grabowska, \emph{{A new basis for
  Hamiltonian SU(2) simulations}},
  \href{https://arxiv.org/abs/2307.11829}{{\ttfamily 2307.11829}}.

\bibitem{gattringer2009quantum}
C.~Gattringer and C.~Lang, \emph{{Quantum Chromodynamics on the Lattice - an
  Introductory Presentation}}, vol.~788, Springer Science \& Business Media
  (2009),
  \href{https://doi.org/10.1007/978-3-642-01850-3}{10.1007/978-3-642-01850-3}.

\bibitem{PhysRevD.11.395}
J.~Kogut and L.~Susskind, \emph{Hamiltonian formulation of {Wilson's} lattice
  gauge theories}, \href{https://doi.org/10.1103/PhysRevD.11.395}{\emph{Phys.
  Rev. D} {\bfseries 11} (1975) 395}.

\bibitem{PhysRevD.91.054506}
E.~Zohar and M.~Burrello, \emph{Formulation of lattice gauge theories for
  quantum simulations},
  \href{https://doi.org/10.1103/PhysRevD.91.054506}{\emph{Phys. Rev. D}
  {\bfseries 91} (2015) 054506}.

\bibitem{Sommer_1994}
{R. Sommer}, \emph{{A new way to set the energy scale in lattice gauge theories
  and its application to the static force and $\alpha_s$ in $\mathrm{SU}(2)$
  Yang-Mills theory}},
  \href{https://doi.org/10.1016/0550-3213(94)90473-1}{\emph{Nucl. Phys. B}
  {\bfseries 411} (1994) }.

\bibitem{Garofalo:2022swa}
M.~Garofalo, T.~Hartung, K.~Jansen, J.~Ostmeyer, S.~Romiti and C.~Urbach,
  \emph{{Defining Canonical Momenta for Discretised SU$(2)$ Gauge Fields}},
  \href{https://doi.org/10.22323/1.430.0040}{\emph{PoS} {\bfseries LATTICE2022}
  (2023) 040} [\href{https://arxiv.org/abs/2210.15547}{{\ttfamily
  2210.15547}}].

\bibitem{Jakobs:2023lpp}
T.~Jakobs, M.~Garofalo, T.~Hartung, K.~Jansen, J.~Ostmeyer, D.~Rolfes et~al.,
  \emph{{Canonical momenta in digitized SU(2) lattice gauge theory: definition
  and free theory}},
  \href{https://doi.org/10.1140/epjc/s10052-023-11829-9}{\emph{Eur. Phys. J. C}
  {\bfseries 83} (2023) 669}
  [\href{https://arxiv.org/abs/2304.02322}{{\ttfamily 2304.02322}}].

\bibitem{Romiti:2023hbd}
S.~Romiti and C.~Urbach, \emph{{Digitizing lattice gauge theories in the
  magnetic basis: reducing the breaking of the fundamental commutation
  relations}},  \href{https://arxiv.org/abs/2311.11928}{{\ttfamily
  2311.11928}}.

\bibitem{PhysRevD.15.1128}
M.~Creutz, \emph{Gauge fixing, the transfer matrix, and confinement on a
  lattice}, \href{https://doi.org/10.1103/PhysRevD.15.1128}{\emph{Phys. Rev. D}
  {\bfseries 15} (1977) 1128}.

\bibitem{PhysRevD.15.1111}
T.~Banks, S.~Raby, L.~Susskind, J.~Kogut, D.R.T.~Jones, P.N.~Scharbach et~al.,
  \emph{Strong-coupling calculations of the hadron spectrum of quantum
  chromodynamics}, \href{https://doi.org/10.1103/PhysRevD.15.1111}{\emph{Phys.
  Rev. D} {\bfseries 15} (1977) 1111}.

\bibitem{PhysRevX.7.041046}
M.C.~Ba\~nuls, K.~Cichy, J.I.~Cirac, K.~Jansen and S.~K\"uhn, \emph{Efficient
  basis formulation for ($1+1$)-dimensional su(2) lattice gauge theory:
  Spectral calculations with matrix product states},
  \href{https://doi.org/10.1103/PhysRevX.7.041046}{\emph{Phys. Rev. X}
  {\bfseries 7} (2017) 041046}.

\bibitem{HUANG1988733}
M.C.~Huang, R.L.~Coldwell and M.W.~Katoot, \emph{A demonstration of the speed
  and accuracy of the biased-selection monte carlo methods in hamiltonian su(2)
  lattice gauge theory},
  \href{https://doi.org/https://doi.org/10.1016/0550-3213(88)90338-0}{\emph{Nuclear
  Physics B} {\bfseries 309} (1988) 733}.

\end{thebibliography}\endgroup

\end{document}